# Hierarchical Deep Learning Model for Degradation Prediction per Look-Ahead Scheduled Battery Usage Profile

Cunzhi Zhao, *Member, IEEE* and Xingpeng Li, *Senior Member, IEEE*

*Abstract*—Batteries can effectively improve the security of energy systems and mitigate climate change by facilitating grid integration of wind and solar power. The installed capacity of battery energy storage system (BESS), mainly the lithium-ion batteries, has increased significantly. However, accurately quantifying battery degradation is challenging but crucial for the economics and reliability of BESS-integrated systems. This paper proposes a hierarchical deep learning-based battery degradation quantification (HDL-BDQ) model to quantify the battery degradation based on scheduled BESS operations. The HDL-BDQ model consists of two deep neural networks that work sequentially. It uses battery operational profiles as input features to accurately estimate the degree of degradation. Additionally, the model outperforms the existing fixed rate or linear rate based degradation models, as well as single-stage deep learning models. The training results demonstrate the high accuracy achieved by the proposed HDL-BDQ model. Moreover, a learning and optimization decoupled algorithm is implemented to strategically leverage the proposed HDL-BDQ model in optimization-based look-ahead scheduling (LAS) problems. Case studies demonstrate the effectiveness of the proposed HDL-BDQ model in LAS of a microgrid testbed.

*Index Terms*— Battery degradation, Battery energy storage system, Deep learning, Energy management system, Machine learning, Microgrid generation resource scheduling, Neural network, Optimization.

## NOMENCLATURE

*Indices:*
| | |
|---|---|
| $g$ | Generator index. |
| $s$ | Battery energy storage system index. |
| $l$ | Load index. |
| $wt$ | Wind turbine index. |
| $pv$ | Photovoltaic index. |

*Sets:*
| | |
|---|---|
| $T$ | Set of time intervals. |
| $G$ | Set of controllable micro generators. |
| $S$ | Set of energy storage systems. |
| $L$ | Set of load. |
| $WT$ | Set of wind turbines. |
| $PV$ | Set of PV systems. |

*Parameters:*

| | |
|---|---|
| $c_g$ | Linear cost for controllable unit $g$. |
| $c_g^{NL}$ | No load cost for controllable unit $g$. |
| $c_g^{SU}$ | Start-up cost for controllable unit $g$. |
| $\Delta T$ | Length of a single dispatch interval. |
| $R_{prcnt}$ | Percentage of the backup power to the total power. |
| $E_s^{Max}$ | Maximum energy capacity of BESS $s$. |
| $c_t^{Buy}$ | Wholesale electricity purchase price in time interval $t$. |
| $c_t^{Sell}$ | Wholesale electricity sell price in time interval $t$. |
| $P_g^{Max}$ | Maximum capacity of generator $g$. |
| $P_g^{Min}$ | Minimum capacity of generator $g$. |
| $P_{Grid}^{Max}$ | Maximum thermal limit of tie-line between main grid and microgrid. |
| $P_g^{Ramp}$ | Ramping limit of diesel generator $g$. |
| $P_s^{Max}$ | Maximum charge/discharge power of BESS $s$. |
| $P_s^{Min}$ | Minimum charge/discharge power of BESS $s$. |
| $\eta_s^{Disc}$ | Discharge efficiency of BESS $s$. |
| $\eta_s^{Char}$ | Charge efficiency of BESS $s$. |

*Variables:*
| | |
|---|---|
| $U_t^{Buy}$ | Status of buying power from main grid in time interval $t$. |
| $U_t^{Sell}$ | Status of selling power to main grid status in time $t$. |
| $U_{s,t}^{Char}$ | Charging status of energy storage system $s$ in time interval $t$. It is 1 if charging status; otherwise 0. |
| $U_{s,t}^{Disc}$ | Discharging status of energy storage system $s$ in time interval $t$. It is 1 if discharging status; otherwise 0. |
| $U_{g,t}$ | Status of generator $g$ in time interval $t$. It is 1 if on status; otherwise 0. |
| $V_{g,t}$ | Startup indicator of Status of generator $g$ in time interval $t$. It is 1 if unit g starts up; otherwise 0. |
| $E_{s,t}$ | Capacity of energy storage system $s$ in time interval $t$. |
| $P_{g,t}$ | Output of generator $g$ in time interval $t$. |
| $P_t^{Buy}$ | Amount of power purchased from main grid power in time interval $t$. |
| $P_t^{Sell}$ | Amount of power sold to main grid power in time interval $t$. |
| $P_{l,t}$ | Demand of the microgrid in time interval $t$. |
| $P_{s,t}^{Disc}$ | Discharging power of energy storage system $s$ at time $t$. |
| $P_{s,t}^{Char}$ | Charging power of energy storage system $s$ at time $t$. |

## I. INTRODUCTION

Renewable energy sources (RES) are widely recognized as a vital component of the future power system due to their numerous advantages over conventional fossil fuels. They offer a clean, sustainable and infinite source of energy that can significantly reduce greenhouse gas emissions and can also contribute to mitigating the effects of climate change [1]. However, the growing usage of RES for power generation has led to stability issues in the system [2]-[3]. To address this challenge, battery energy storage systems (BESS) are being

Cunzhi Zhao and Xingpeng Li are with the Department of Electrical and Computer Engineering, University of Houston, Houston, TX, 77204, USA (e-mail: czhao20@uh.edu; Xingpeng.Li@asu.edu).

This work is sponsored by the Grants to Enhance and Advance Research (GEAR) program by Division of Research at the University of Houston.

adopted as a solution, BESS can effectively balance the variable and uncertain nature of RES by storing excess energy generated during periods of high generation and releasing it during periods of low generation [4]-[5]. This enables a more reliable and stable integration of renewable energy into the power system. Moreover, BESS can provide important ancillary services like frequency regulation, voltage control, and peak shaving, which enhance the stability and efficiency of the overall power system [6].

Numerous studies have consistently demonstrated the effective integration of BESS into both bulk power systems and microgrids. Notably, a compelling illustration of BESS's positive impact is the offshore system outlined in [8], which demonstrably reduces carbon emissions. Furthermore, the findings in papers [9]-[11] underscore the substantial advantages of incorporating BESS into power systems. Papers [12]-[13] showcase how microgrids, when equipped with BESS, can seamlessly support the main grid. In addition to these examples, several models have been proposed to incorporate BESS, aiming to address and mitigate fluctuations induced by renewable energy sources, as elucidated in [14]-[16]. To summarize, the deployment of BESS emerges as a pivotal factor for the successful integration of renewable energy into the power system. Beyond enhancing system stability and efficiency, the incorporation of BESS lays the foundation for a cleaner and more sustainable energy future.

The main component of BESS currently on the market is lithium-ion batteries (LIB). However, the characteristics of these batteries cause them to degrade over time, which can negatively affect their performance and efficiency [17]. LIB degrade over time due to a variety of factors such as repeated charge and discharge cycles, high temperatures, and high charging/discharging rate. The repeated movement of ions within the battery can also cause physical damage to the electrodes and electrolyte, leading to a reduction in capacity and efficiency over time. Moreover, the accumulation of impurities can further contribute to degradation [18]. All these factors can result in a decline in the overall lifespan and performance of the lithium-ion battery. LIB are chemical battery, and the degradation is mainly caused by the loss of Li-ions, electrolyte, and the increase of internal resistance. Factors such as ambient temperature, charging/discharging rate, state of charge (SOC), state of health (SOH), and depth of discharge (DOD) are the key factors that can affect the degradation during battery cycling [19]-[20]. Furthermore, the internal resistance [21] and internal temperature [22] are highly related to battery degradation and safety. High temperatures can promote side reactions, such as electrolyte decomposition or electrode corrosion, which can degrade the battery's components over time. This internal chemical reaction is hard to model [6]. The internal resistance of a battery affects its efficiency and performance. Increasing internal resistance puts stress on the battery during operation, which can lead to physical damages or changes in the structure of the electrodes and electrolyte, accelerating degradation [21]. However, accurately assessing the internal state of the battery remains challenging, especially as batteries are becoming increasingly important as energy storage systems in both microgrids and bulk power systems. Predicting battery degradation is therefore a challenging difficult task, particularly when BESS operates in various conditions and environments [23]. Therefore, an accurate and efficient battery degradation model is urgent for the energy management system especially as a significant number of BESS are being installed into power systems.

Several previous studies have developed some battery degradation models, but these models have limitations in accurately predicting degradation under different scenarios. For instance, a DOD based quadratic function is applied in [24] to estimate the capacity fade for each cycle. The maximum theoretically achievable number of cycles is also involved within the degradation prediction. However, the proposed degradation model in [24] only considers SOC and DOD as the variables and neglects the effect of other battery degradation factors (BDF) which could result in a huge error in battery degradation prediction under different scenarios. Similarly, various DOD-based models proposed in [25]-[29] also neglect other significant degradation factors. A constant rate of battery degradation is proposed in [30]. The degradation cost is linear with the power output of the BESS. This may approximate the battery degradation in the short term, but it will lead to huge error of degradation prediction in the long run since the degradation is highly non-linear and related to multiple battery degradation factors. The linear assumptions of battery degradation in [31]-[33] simplify the problem and reduce computational complexity but may not accurately predict degradation. The heuristic battery degradation models can be categorized as either linear degradation or DOD based models. However, a quality examination of those two models conducted in [34] indicates that they cannot accurately predict the degradation values caused by different ambient temperatures, and the degradation prediction error is high for both models. Most heuristic models, including DOD-based and linear degradation models, have limitations because they oversimplify battery degradation. They often rely solely on metrics like the depth of discharge or other single degradation features. These models lack the necessary detail to fully capture the complex factors that affect battery health, such as SOC, ambient temperature, charge/discharge rate, and SOH. As a result, these models often lack precision in predicting battery degradation, especially over the long term, because they neglect these crucial parameters. Additionally, the static nature of DOD-based models makes them challenging to adapt to dynamic operating conditions. This limitation means they may not accurately reflect the complexities of real-world scenarios, especially in systems with varied and unpredictable usage patterns.

A pattern-driven degradation model is designed in [35] to predict the remaining useful life cycles. It is a learning based model that leverages the historical battery data. However, incorporating this model into the look-ahead scheduling (LAS) problems can be challenging due to computational complexity. Similarly, a battery degradation matrix model [36] is also impractical to be integrated into the LAS problems. A data driven model based on historical data of starting SOC and ending SOC is proposed in [37] to predict the capacity loss. The model has a high accuracy based on its dataset, but it cannot accurately predict the degradation level when the BESS operating scenarios are out of its dataset. Similarly, a temperature based battery degradation model is proposed in [38]; a quadratic equation is formed based on the collected data [39]; however, the data that applied to those data driven models are limited and thus the model fails to consider other important degradation factors. Overall, while various models have been proposed to predict battery degradation, their limitations and assumptions should be carefully considered when applied in different scenarios.

In summary, all the aforementioned battery degradation models can be summarized as follows:

1) Linear Degradation Model: A linear degradation cost based on the power usage or energy consumption is applied in the battery degradation model. This may lead to a huge error on the battery degradation under different operating conditions [40].
2) DOD-based Model: The battery degradation is calculated based on the DOD per cycle, which omits other important battery degradation features.
3) Learning-based/Data-Driven Model: The battery degradation is predicted by a learning model which is non-linear. The complexity of the model makes it hard to incorporate with the LAS optimization problems [41].

A fully connected neural network (NN) [41] based battery degradation model and an iterative solving algorithm are proposed to address the above-mentioned gaps in [35]-[40]. This model exhibits robust performance across diverse scenarios, boasting a 94.5% accuracy in degradation prediction within a 15% error tolerance. Notably, its input parameters encompass ambient temperature, charging/discharging rate, SOC, DOD, and SOH. However, the omission of internal factors like internal temperature and resistance in the previous NN model neglects key influencers on battery degradation such as internal temperature and internal resistance. In response to this limitation, we introduce a pioneering solution: the hierarchical deep learning-based battery degradation quantification (HDL-BDQ) model, illustrated in the left segment of Fig. 1. Compared to references [35]-[41], the proposed HDL-BDQ model provides a more holistic view of battery degradation by incorporating a broader spectrum of degradation features and addressing the look-ahead scheduling problem in a more integrated manner. The look-ahead problem complicates the acquisition of certain degradation features because they cannot be directly obtained from the immediate results of the look-ahead process. By effectively predicting these features through our proposed DNN framework, we ensure that the look-ahead scheduling is informed by a more complete and accurate representation of battery degradation, ultimately leading to better decision-making in battery management systems.

This innovative HDL-BDQ model aims to enhance the accuracy of battery degradation training by deploying two sequential deep neural networks (DNNs): DNN for Unobtainable Battery Degradation Features (DNN-UBDF) and DNN for Battery Degradation Prediction (DNN-BDP). The DNN-UBDF specializes in predicting elusive battery degradation features which are challenging to acquire directly from LAS. These predicted features then serve as inputs for the DNN-BDP, enabling it to precisely quantify the corresponding battery degradation based on the scheduled usage profile. This hierarchical approach ensures a more comprehensive understanding of the intricate factors influencing battery health, leading to improved accuracy and reliability in degradation predictions.

The HDL-BDQ models are specifically designed to measure battery degradation. Nonetheless, solving a DNN embedded LAS problem directly poses challenges due to the inherently non-linear and non-convex nature of the DNN model. To overcome this obstacle, we introduce a learning and optimization decoupled (LOD) algorithm tailored for efficiently tackling the embedded LAS optimization problem posed by HDL-BDQ. In Fig. 1's right segment, the LOD algorithm undertakes iterative solutions to the HDL-BDQ model's embedded LAS, decoupling it from the battery degradation calculation and LAS optimization problems. Each LOD algorithm iteration updates constraints on Battery Energy Storage System (BESS) operation to curtail usage, thereby minimizing both battery degradation and associated costs in subsequent iterations. However, limiting BESS output raises microgrid operation costs. The LOD algorithm's overarching objective is to determine the optimal balance between battery degradation and microgrid operation costs. Throughout the LOD algorithm iterations, the algorithm maintains a record of the total cost and strives to find the optimal solution for the HDL-BDQ embedded LAS problem at the vertex point. This ensures an optimal solution concerning the vertex point of the combined total cost. The potential users of the proposed solution could include utilities, microgrid operators, or third-party BESS owners, depending on who owns/operates the BESS. The main contributions of this paper can be summarized as follows:

- Sequential DNN Construction:
  Two sequential DNNs are constructed for the HDL-BDQ model to quantify the battery degradation based on the given operational profile.
- DNN Model Scalability:
  Multiple DNN architectures are proposed for both DNN-UBDF and DNN-BDP to compete for the best training model that has the highest accuracy of battery degradation prediction.
- Superior Performance of HDL-BDQ:
  The proposed HDL-BDQ model demonstrates exceptional performance, outperforming other existing models in terms of battery degradation prediction accuracy.
- LOD Algorithm Implementation:
  The implementation of the LOD algorithm skillfully addresses the computational complexity of the learning model-embedded look-ahead optimal energy scheduling problem posed by integrating HDL-BDQ into LAS.
- Performance Demonstration in a Microgrid:

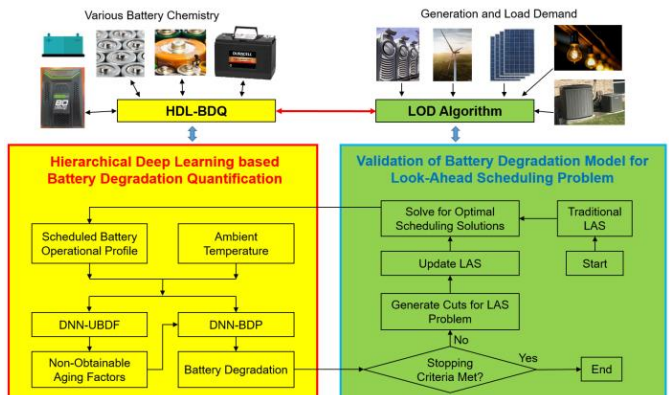

Fig. 1. HDL-BDQ and LOD algorithm.

Illustration of the HDL-BDQ model's outstanding performance through the application of the LOD algorithm in a microgrid system.

The rest of the paper is organized as follows. Section II describes the HDL-BDQ model. Section III presents the training algorithm and setups of the proposed DNN training models. Section IV presents the LOD algorithm. Section V discusses the case study with a microgrid testbed and Section VI concludes the paper.

## II. HIERARCHICAL DEEP LEARNING BASED BATTERY DEGRADATION QUANTIFICATION MODEL

The proposed HDL-BDQ model utilizes two sequential deep neural networks to adaptively quantify the battery degradation associated with the scheduled battery operational profiles over multiple time intervals. Several factors affect battery degradation during a (dis)charge cycle, as listed in Table I including (i) ambient temperature (Temp), (ii) current battery energy capacity or SOH, (iii) SOC level, (iv) DOD level, (v) (dis)charge rate (C Rate), (vi) battery internal temperature (IT), (vii) battery internal resistance and (viii) equivalent life cycle numbers (ELCN) which varies per other degradation features. However, battery internal temperature, battery internal resistance and equivalent life cycle numbers cannot be obtained explicitly in advance. To address this issue, we have developed a DNN-UBDF model that uses the first five features to predict these unknown factors. Subsequently, the output of DNN-UBDF model, along with the other features, will be used as inputs to the trained DNN-BDP model to predict and quantify battery degradation. By comparing the training results of the proposed DNN models, we can identify the most efficient and accurate combination for the proposed HDL-BDQ model.

Table I Battery degradation factors

| Features | Obtainable/Available from LAS |
|---|---|
| State of Charge | Yes |
| Depth of Discharge | Yes |
| Ambient Temperature | Yes |
| Charge/Discharge Rate | Yes |
| Current Capacity | Yes |
| Internal Temperature | No |
| Internal Resistance | No |
| Equivalent Life Cycle Numbers | No |

The HDL-BDQ proposed in this paper consists of two deep neural networks: the first DNN model (DNN-UBDF) will predict the battery internal parameters that cannot be calculated or measured directly per the scheduled battery operational profile in advance. The outputs of DNN-UBDF, which are the predicted battery parameters, will then be used as additional input features for DNN-BDP, along with the scheduled battery operational profile, to quantify the corresponding battery degradation in the same look-ahead time period.

### A. DNN-Unobtainable Battery Degradation Features

As mentioned in the introduction section above, some highly influential battery degradation factors such as battery internal temperature and internal resistance are hard to predict or measure in the look-ahead scheduling. Thus, DNN-UBDF is proposed to capture the non-linearity between the available variables and unknown features that are highly correlated with battery degradation. IT, IR and LCN are hard to measure or calculate based on the available information as shown in Table I. To address this issue, several potential DNN models are proposed and listed in Table II. DNN-UBDF is modeled using a fully connected neural network. The deep neural network takes five critical factors (ambient temperature, C rate, SOC, DOD, and SOH) as inputs to predict the value of IT, IR and/or ELCN. A dynamic learning rate, which decreases automatically after a specified number of epochs, is employed during the training process to enhance the results. The trained network has a 5-neuron input layer, 20-neuron first hidden layer, and a 10-neuron second hidden layer. The number of the output neurons depends on the number of outputs in the proposed model in Table II. The activation functions used are rectified linear unit (relu) for the hidden layers and "linear" for the output layer, as shown in Fig. 2 plotted with Lucidchart [42].

Table II Potential models for DNN-UBDF

| Model # | Inputs | Outputs |
|---|---|---|
| 1 | SOC, DOD, Temp, C Rate, SOH | IT |
| 2 | SOC, DOD, Temp, C Rate, SOH | IR |
| 3 | SOC, DOD, Temp, C Rate, SOH | IT, IR |
| 4 | SOC, DOD, Temp, C Rate, SOH | IT, ELCN |
| 5 | SOC, DOD, Temp, C Rate, SOH | IR, ELCN |
| 6 | SOC, DOD, Temp, C Rate, SOH | IT, IR, ELCN |

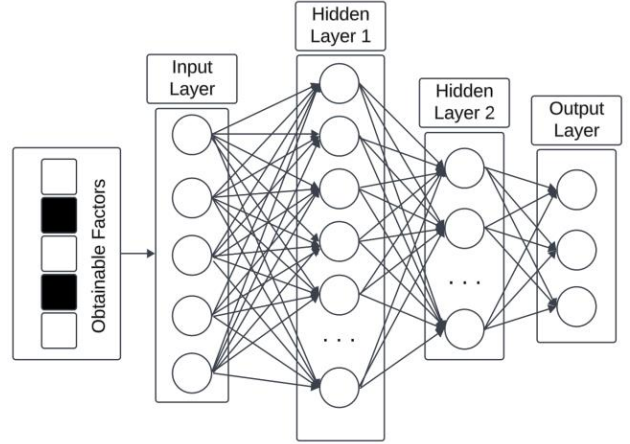

Fig. 2. Structure of the sample neural network model for DNN-UBDF.

### B. DNN-Battery Degradation Prediction

DNN-BDP is proposed to predict the battery degradation value with the inputs of major battery degradation factors, including the outputs of DNN-UBDF. As these factors are highly correlated to battery degradation, including them as inputs can significantly improve the training results. Similar to DNN-UBDF, we proposed several potential DNN models for DNN-BDP as shown in Table III: several different combinations of critical battery degradation factors are utilized separately to predict the battery degradation value in terms of percentage with respect to SOH for a given cycle. The structure of DNN-BDP models is illustrated in Fig. 3. The number of input neurons depends on the number of input features of the proposed DNN-BDP models as listed in Table III. The model includes a first hidden layer with 20 neurons, a second hidden layer with 10 neurons, and a single neuron in the output layer specifically for degradation prediction.

While the DNN-UBDF models in Table II share the same inputs, they are distinguished by their output configurations. Each model is specifically trained to predict one or more battery degradation features, ensuring that the predicted outputs are tailored to the specific needs of the subsequent DNN-BDP models. We systematically explored these combinations using grid search to identify the best-performing configuration for the HDL-BDQ model. This approach allows for a more accurate modeling of battery degradation, addressing the gaps identified in the literature and improving prediction accuracy in real-world applications.

Table III Potential models for DNN-BDP

| Model # | Inputs | Outputs |
|---|---|---|
| 1 | IT, ELCN | Degradation |
| 2 | IR, ELCN | Degradation |
| 3 | SOC, DOD, Temp, C Rate, IT | Degradation |
| 4 | SOC, DOD, Temp, C Rate, IR | Degradation |
| 5 | SOC, DOD, Temp, C Rate, IT, ELCN | Degradation |
| 6 | SOC, DOD, Temp, C Rate, IR, ELCN | Degradation |
| 7 | SOC, DOD, Temp, C Rate, IT, SOH | Degradation |
| 8 | SOC, DOD, Temp, C Rate, IR, SOH | Degradation |
| 9 | SOC, DOD, Temp, C Rate, IT, SOH, ELCN | Degradation |
| 10 | SOC, DOD, Temp, C Rate, IR, SOH, ELCN | Degradation |

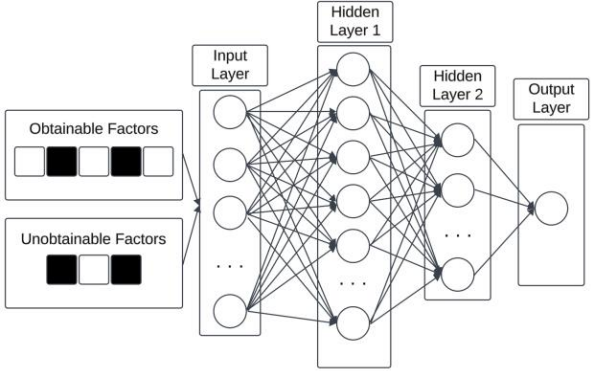

Fig. 3. Structure of the sample neural network model for DNN-BDP.

## III. TRAINING PROCESS AND RESULT COMPARISON OF THE PROPOSED HDL-BDQ MODEL

### A. Input Data

The training of the deep neural networks requires large amount of data. To this end, we utilize MATLAB Simulink [43] to conduct battery aging tests by implementing a battery model. We employ a battery cycle generator to simulate charging and discharging cycles at preset rates, enabling adjustment of SOC and simulation of various battery types, conditions, and operating profiles, as well as ambient temperature effects and internal resistance. Battery aging tests are conducted at different initial SOC and DOD levels, under different ambient temperatures and charging/discharging rates. The aging test model integrates the battery model and cycling generator within Simulink. By adjusting the related battery operating parameters, the variables can be modified for each aging test. For example, the SOC variable can be adjusted in the battery settings. The battery model includes several lithium battery types, and the battery parameters can be customized using the manufacturer's datasheet if a new battery type needs to be introduced. Additionally, the simulation can simulate ambient temperature effects. To present the realistic representations, multiple battery aging tests based on the various combination of battery degradation features have been simulated. There are 35 groups of battery aging tests collected under various combinations of SOC, DOD, Temp and C rate as shown in Table IV.

Table IV Number of battery aging tests under various conditions.

| DOD | C Rate | | | |
|---|---|---|---|---|
| | 1 C | 1.5 C | 2 C | 5 C |
| 20% | 2 | 2 | 2 | 2 |
| 60% | 2 | 2 | 2 | 2 |
| 80% | 2 | 2 | 2 | 2 |
| 100% | 2 | 3 | 3 | 3 |

The collected data from the battery aging tests, including internal resistance, internal temperature, ELCN, and energy capacity comprise the training dataset for DNN-UBDF and DNN-BDP, as previously discussed. The degradation per cycle is calculated as the percentage change in SOH from the beginning to the end of that cycle. The SOH data is collected at the end of each charging/discharging cycle when the battery attains its preset SOC value. Each cycle represents the process of discharging a battery from a specific SOC to a lower SOC and then recharging it back to the original SOC. It is typically considered as the end of a battery's life when its capacity degrades to 80% of the maximum rated capacity [44]. Thus, the ELCN here represents the cycle numbers when the capacity of battery reaches 80% of the maximum capacity. The charging/discharging rate is referred to as C rate, which measures the speed at which a battery is charged or discharged as defined in (1). $I_{BESS}$ represents the charging/discharging current of the BESS and $E_{Max}$ denotes the rated capacity of BESS.

$$C\ rate = \frac{I_{BESS}}{E_{Max}} \quad (1)$$

### B. Pre-training Setup

The battery degradation data collected from the Simulink need to be normalized in order to improve the training efficiency and accuracy before sending to the training model. The original data include the SOC, DOD, Temp, C Rate, IT, IR, ELCN and SOH. To increase the training efficiency, SOC, DOD, Temp, IT, IR and ELCN are normalized. The normalization used in [45] is applied in the data pre-processing in this work and shown in (2):

$$\widehat{x_k} = \frac{x_k - x_{min}}{x_{max} - x_{min}} \quad (2)$$

where $x_{max}$ and $x_{min}$ represent the largest value and the lowest value of $x_k$ respectively. The degradation value is calculated as the difference of SOH between cycles from the raw data without any filtering or pre-processing. The input data will be split into two parts, 80% as the training dataset and 20% as the testing dataset.

During training, the deep neural network is trained with the mini-batch gradient descent strategy. The deep neural network's performance is evaluated based on its ability to accurately predict its outputs, which is measured using the mean squared error (MSE) between the predicted and actual values. MSE is calculated by taking the average of the square of the difference between the actual and predicted values across all the training data points and is used as the loss function during the training process as defined in (3). (4) and (5) are used to assess the accuracy of the neural network model on the testing dataset, where $y^{predict}$ in (4) denotes the predicted value and $y^{actual}$ represents the actual value. We evaluate predictions by applying thresholds of 5%, 10%, 15%, and 20%. If the prediction error falls below the chosen threshold, the prediction is deemed accurate and converted to a binary output (1); other-wise, it is marked as inaccurate (0). This threshold-based conversion is consistently applied to both the DNN-UBDF and DNN-BDP models. For models with multiple outputs, such as Model 6 from Table II, each output is independently evaluated using the error tolerance criteria. Specifically, every predicted value is compared with its corresponding actual value using the error formula. For example, if a model generates 300 outputs, with each output containing 3 feature predictions, a total of 900 predictions are evaluated. Each of these predictions is assessed against the error threshold to determine its accuracy (binary 1 for within the threshold, binary 0 for outside the threshold). The overall accuracy is then calculated as the ratio of correct predictions to the total number of predictions with (5).

$$MSE = \frac{1}{n}\sum_{i=1}^{n}(y_i - \tilde{y}_i)^2 \quad (3)$$

$$error = \frac{|y^{predict} - y^{actual}|}{y^{actual}} \quad (4)$$

$$Accuracy = \frac{Number\ of\ accurate\ predictions}{Number\ of\ total\ predictions} \quad (5)$$

### C. Benchmark Models

a. **Benchmark NNBD**:
The neural network based battery degradation (NNBD) model [41] that uses SOC, DOD, C rate, temperature and SOH as the inputs to predict the battery degradation is served as the benchmark model for this paper. As discussed in the introduction section, the accuracy of this model is 94.5% with an error tolerance of 15%. It is worth mentioning that its training dataset was pre-processed in addition to normalization to obtain this accuracy.

b. **Benchmark NNBD2**:
Another benchmark is proposed with an extra hidden layer included into the first benchmark model. Since the proposed HDL-BDQ model utilizes two DNNs to predict the battery degradation. It is possible that a single stage DNN model with extra hidden layer as shown in Fig. 4 may also achieve similar performance.

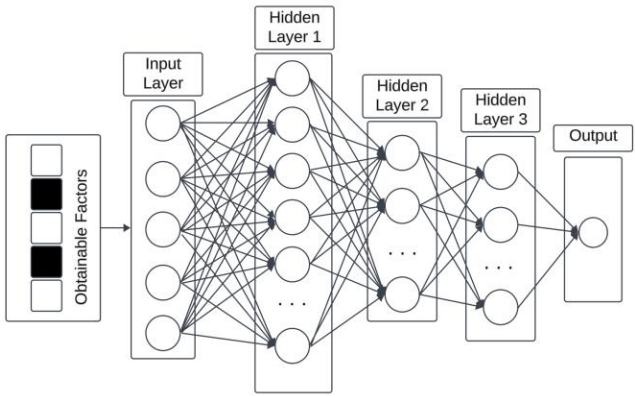

Fig. 4. Structure of the proposed Benchmark 2 model.

## IV. THE PROPOSED HDL-BDQ EMBEDDED ENERGY SCHEDULING FOR MICROGRID SYSTEMS

In this section, the traditional microgrid LAS model is presented, as well as the proposed HDL-BDQ embedded LAS model, followed by the implemented LOD algorithm.

### A. Traditional Look-Ahead Scheduling Problem

The objective of the traditional LAS model is to minimize the total operating cost of the microgrid as defined in (5). The balance of power within the microgrid, taking into account the contribution from controllable generators, renewable energy sources, power exchange with the main grid, battery output, and electrical demand, is described in (6). Constraint (7) sets limits on the power outputs of controllable units, such as diesel generators. The ramp-up and ramp-down limitations are imposed by constraints (8) and (9) respectively. Equation (10) determines the status of power exchange between the microgrid and main grid, either as a buyer, seller, or idle. Constraints (11) and (12) enforce the thermal limits of the tie-line. Equation (13) restricts the battery to be in either charging mode, discharging mode, or idle mode. Constraints (14) and (15) limit the charging and discharging power of the battery. As shown in (16), the battery's SOC can be calculated based on its current energy storage. Equation (17) calculates the energy stored in the battery at each time interval, and (18) requires the final SOC level of the battery to match its initial value. Constraint (19) ensures that the microgrid has adequate backup power available to deal with power outage events.

*Objective function*:
$$Min\ f^{Cost} = \sum_{t \in T}\sum_{g \in G}\left( (P_{g,t}c_{g,t} + U_{g,t}c_g^{NL} + V_{g,t}c_g^{SU}) + P_t^{Buy}c_t^{Buy} - P_t^{Sell}c_t^{Sell} \right) \quad (5)$$

*Constraints*:
$$P_t^{Buy} + \sum_{g \in G}P_{g,t} + \sum_{wt \in WT}P_{wt,t} + \sum_{pv \in PV}P_{pv,t} + \sum_{s \in S}P_{s,t}^{Disc} = P_t^{Sell} + \sum_{l \in L}P_{l,t} + \sum_{s \in S}P_{s,t}^{Char} \quad (6)$$

$$P_g^{Min} \leq P_{g,t} \leq P_g^{Max}, \forall g,t \quad (7)$$

$$P_{g,t+1} - P_{g,t} \leq \Delta T \cdot P_g^{Ramp}, \forall g,t \quad (8)$$

$$P_{g,t} - P_{g,t+1} \leq \Delta T \cdot P_g^{Ramp}, \forall g,t \quad (9)$$

$$U_t^{Buy} + U_t^{Sell} \leq 1, \forall t \quad (10)$$

$$0 \leq P_t^{Buy} \leq U_t^{Buy} \cdot P_{Grid}^{Max}, \forall t \quad (11)$$

$$0 \leq P_t^{Sell} \leq U_t^{Sell} \cdot P_{Grid}^{Max}, \forall t \quad (12)$$

$$U_{s,t}^{Disc} + U_{s,t}^{Char} \leq 1, \forall s,t \quad (13)$$

$$U_{s,t}^{Char} \cdot P_s^{Min} \leq P_{s,t}^{Char} \leq U_{s,t}^{Char} \cdot P_s^{Max}, \forall s,t \quad (14)$$

$$U_{s,t}^{Disc} \cdot P_s^{Min} \leq P_{s,t}^{Disc} \leq U_{s,t}^{Disc} \cdot P_s^{Max}, \forall s,t \quad (15)$$

$$SOC_t^s = E_{s,t}/E_s^{Max}, \forall s,t \quad (16)$$

$$E_{s,t} - E_{s,t-1} + \Delta T \cdot \left(P_{s,t-1}^{Disc}/\eta_s^{Disc} - P_{s,t-1}^{Char}\eta_s^{Char}\right) = 0, \forall s,t \quad (17)$$

$$E_{s,t=24} = E_s^{Initial}, \forall s \quad (18)$$

$$P_{Grid}^{Max} - P_t^{Buy} + P_t^{Sell} + \sum_{g \in G}(P_g^{Max} - P_{g,t})$$
$$\geq R_{prcnt}\left(\sum_{l \in S_L} P_{l,t}\right), \forall t \quad (19)$$

### B. HDL-BDQ Embedded LAS

The traditional LAS model described in Section IV.A provides an initial solution without considering the impact of battery degradation. This method optimizes the operational profiles for the BESS, controllable generators, and tie-line exchange power, but assumes an ideal BESS without degradation, leading to a zero equivalent battery degradation cost. This can lead to economic loss in the long term due to accelerated aging and costly replacement of BESS. To overcome the limitation, the HDL-BDQ model is designed to accurately account for the BESS degradation in the optimal scheduling. The proposed HDL-BDQ model requires the SOC level from the BESS operation profile as input. The DOD level is calculated as the absolute value of the difference between the SOC levels at time intervals $t$ and $t-1$, as shown in (20). The C rate is calculated by (21). The SOH level of the battery is assumed to be available before the LAS. The input vector $\overline{x}_t^{DNN-UBDF}$, as shown in (22), can be formed by combining these values and then fed into the trained DNN-UBDF model to predict the related unobtainable battery degradation features (24). Then, DNN-BDP model along with the input vector of $\overline{x}_t^{DNN-BDP}$ (23) will be able to obtain the total battery degradation over the LAS time horizon as shown in (25). Note that the input battery degradation features for $\overline{x}_t^{DNN-UBDF}$ and $\overline{x}_t^{DNN-BDP}$ are optimized and determined by comparing the overall performance of the proposed DNN models in Table II and Table III.

$$\Delta DOD_t = |SOC_t - SOC_{t-1}| \quad (20)$$
$$C_t^{Rate} = \Delta DOD_t / \Delta T \quad (21)$$
$$\overline{x}_t^{DNN-UBDF} = (BDF) \quad (22)$$
$$\overline{x}_t^{DNN-BDP} = (BDF\ \&\ UBDF) \quad (23)$$
$$UBDF = \sum_{t \in S_T} f^{DNN-UBDF}(\overline{x}_t^{DNN-UBDF}) \quad (24)$$
$$Degradation = \sum_{t \in S_T} f^{DNN-BDP}(\overline{x}_t^{DNN-BDP})SOH \quad (25)$$

A Cycle Based Usage Processing (CBUP) method [41] is applied here to address the inconsistency of the dynamic LAS scheduling and the fixed cycle-based HDL-BDQ model. Since the BESS is not scheduled with fixed cycles and constant charge/discharge rates per LAS solutions, CBUP can help approximate and aggregate the BESS hourly operational profile into fixed cycles that can then be used by the proposed HDL-BDQ model to estimate the associated battery degradation. As a result, (24)-(25) are replaced by (26)-(27) while $\overline{x}_c^{DNN-UBDF}$ and $\overline{x}_c^{DNN-BDP}$ represent the input vectors for the aggregated cycles and $AC$ represents the set of feasible aggregated cycles.

$$UBDF = \sum_{t \in AC} f^{DNN-UBDF}(\overline{x}_c^{DNN-UBDF}) \quad (26)$$
$$Degradation = \sum_{t \in AC} f^{DNN-BDP}(\overline{x}_c^{DNN-BDP})SOH \quad (27)$$

The objective function of LAS needs to be revised to account for the cost of battery degradation in the HDL-BDQ embedded LAS model. This updated objective function is presented in (28). This equation comprises of $f^{cost}$, defined in (4), and $f^{BESS}$, which represents the battery degradation cost estimated by the proposed HDL-BDQ model. The calculation of $f^{BESS}$ can be done in (29), which involves the capital investment cost of BESS $c_{BESS}^{Capital}$, the salvage value at the end of life $c_{BESS}^{SV}$, the state of health value at the end of life $SOH_{EOL}$, and the percentage of battery degradation.

$$f = f^{Cost} + f^{BESS} \quad (28)$$
$$f^{BESS} = \frac{c_{BESS}^{Capital} - c_{BESS}^{SV}}{1 - SOH_{EOL}} Degradation \quad (29)$$

Thus, the proposed HDL-BDQ embedded LAS model can be represented by (5)-(19), (20)-(23), and (26)-(29).

### C. Learning and Optimization Decoupled Algorithm

The proposed hierarchical HDL-BDQ embedded LAS problem, as discussed before, is difficult to solve directly due to the non-linear and non-convex nature of the multi-layer HDL-BDQ model. To overcome this challenge, the Learning and Optimization Decoupled Algorithm from our previous work [41] is enhanced and customized in this work to simplify the complex multi-layer HDL-BDQ embedded LAS problem. This algorithm works iteratively through five steps as shown in Fig. 5.

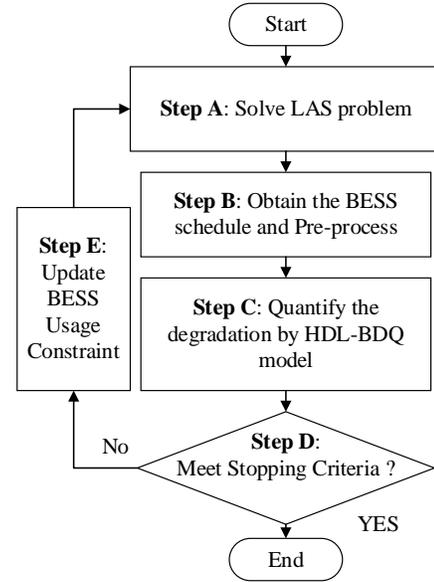

Fig. 5. LOD Algorithm flowchart.

• Step A solves the LAS problem with updated constraints by restricting the usage of BESS as determined from Step E. For the first iteration, there are no restrictions on BESS usage.

• Step B retrieves the scheduled BESS operating profile which is then processed with the proposed CBUP method.

• Step C utilizes the pre-trained two sequential DNN models to estimate BESS degradation and calculate the equivalent battery degradation cost.

• Step D checks if the solutions meet the designated stopping criteria. If yes, the iteration process stops and the solution is reported, otherwise it proceeds to Step E.

• Step E adjusts the limits of the BESS operating constraints, which can decrease the degradation cost. The updated constraints are then passed on to Step A for inclusion of LAS in the next iteration.

The objective of Step E is to create updated restrictions that control the battery's output to minimize overall battery degradation. The updated constraints will further limit the total energy consumption in which the BESS can operate after each iteration. As shown in (30) [41], it aims to decrease the aggregated battery output power over a 24-hour period by mandating it to be lower than the previous iteration. In each iteration, constraint (30) will be updated by utilizing the scheduled battery operational profile from the previous iteration and the preset parameter α, which determines the reduction of total energy consumption per iteration. $P_{BatteryTotal}^{Pervious\_SCUC}$ represents the total charging and discharging power based on the BESS scheduling from previous SCUC solution, which is calculated by the left side of (29) based on the optimized schedule from previous iteration.

$$\sum_{t \in T}(P_{s,t}^{Char} + P_{s,t}^{Disc})\Delta T \leq (1-\alpha) * P_{BatteryTotal}^{Pervious\_SCUC} \quad (30)$$

### D. Stopping Criteria

The objective of Step A in LOD algorithm is to find the optimal solution for the LAS model without considering the impact of battery degradation. The LOD algorithm method ensures that a solution is guaranteed to be optimal. The first iteration finds the optimal outcome when battery degradation is not taken into account. Subsequent iterations will restrict the operation of the BESS to minimize battery degradation. As the iteration increases, the battery degradation will decline, but the operation cost will increase. The optimal solution is the point at which the total cost curve reaches its minimum. A termination criterion is set to improve the solving efficiency. The termination criteria for the iterations are set using a designed stopping rule, with the objective of identifying the best solution. Specifically, if the total cost of iteration $i$ is the lowest point between iteration i-10 and i+10, the iterative process will terminate at iteration i+10, and the solution of iteration $i$ will be with the lowest total cost for the HDL-BDQ embedded LAS problem. Thus, the LOD algorithm is a look-back evaluation method: when at iteration $i$, we examine whether the solution for iteration $i$-10 is the optimal solution. If the optimal solution is within the very first 10 iterations, then a smaller value of α is suggested to extend the total iteration number and obtain precise solutions.

### E. Benchmark Models for HDL-BDQ Embedded LAS

We proposed two benchmark models to evaluate and demonstrate the effectiveness and performance of the proposed HDL-BDQ embedded LAS model and LOD algorithm. Details of two benchmark models are presented below:
1. *Traditional LAS*: The traditional benchmark model is established without incorporating HDL-BDQ model, and the objective is to optimize the system's operating cost by maximizing the utilization of BESS.
2. *Linear Battery Degradation Cost (BDC) Model*: The Linear BDC Model operates under the assumption that the cost of battery degradation increases linearly with the energy consumption of the BESS. This is reflected in (31) where the rate of degradation, $c_{BESS}$, is a fixed constant cost per kWh.

$$f^{BESS} = c_{BESS} \sum_{s \in S_S, t \in S_T} P_{s,t}^{Char/Disc} * \Delta T \quad (31)$$

## V. CASE STUDIES

### A. DNN-UBDF Training Results

The testing results of proposed DNN-UBDF models are presented in Table V based on the testing dataset. From this table, we can observe that the training accuracies are similar at 15% error tolerance level between DNN-UBDF Model 1 and Model 2, as well as Model 4 and Model 5. However, the accuracies of the various models vary at different error tolerance levels. For instance, while DNN-UBDF Model 1 outperforms Model 5 at the 10% error tolerance level, Model 5 surpasses Model 1 at the 15% tolerance level. The main objective is to have a high training accuracy HDL-BDQ model in this paper. Thus, the selection of DNN-UBDF is based on the performance of the proposed DNN-BDP models. DNN-BDP can determine the inputs that lead to the highest accuracy of the ultimate battery prediction model.

Table V Testing results of proposed models for DNN-UBDF

| Error Tolerance | 5% | 10% | 15% | 20% |
| --- | --- | --- | --- | --- |
| DNN-UBDF Model 1 | 45.25% | 77.33% | 88.63% | 89.74% |
| DNN-UBDF Model 2 | 48.66% | 78.37% | 89.21% | 90.59% |
| DNN-UBDF Model 3 | 37.96% | 69.14% | 85.47% | 87.19% |
| DNN-UBDF Model 4 | 48.82% | 73.86% | 88.58% | 91.23% |
| DNN-UBDF Model 5 | 51.05% | 75.12% | 89.90% | 90.11% |
| DNN-UBDF Model 6 | 39.78% | 65.61% | 81.31% | 83.74% |

### B. DNN-BDP Training Results

Based on the results shown in Table VI, it is evident that the DNN-BDP Model 10 outperforms all other models across the tolerance levels. Thus, it can be concluded that, given the training dataset, Model 10 which takes SOC, DOD, Temp, C Rate, IR, SOH and ELCN as inputs, is the best model for DNN-BDP in battery degradation prediction. Model 10 includes the most input features among the models besides Model 9. Model 1 and Model 2 are the simplest model that consists only two input features and perform well, but the accuracy at 5% tolerance level is low comparing to Model 10. Conversely, Model 7 and Model 8 yield very low prediction accuracy. Interestingly, the prediction accuracy of Model 3 and Model 4 are quite high. The only difference is the SOH is added in the input features for Model 7 and Model 8. Moreover, when we compare Model 8 and Model 10, the huge accuracy difference shows the importance of ELCN in battery degradation prediction. Further analysis shows that even-numbered models, which include IR as an input variable, perform better than odd-numbered models, which include IT. This suggests that IR is a better indicator of degradation than IT, at least on the dataset that has been examined in this work. Fig. 6 shows the results in the training process of the Model 10 that becomes stable after 200 epochs.

Table VI Testing results of proposed models for DNN-BDP

| Error Tolerance | 5% | 10% | 15% | 20% |
| --- | --- | --- | --- | --- |
| DNN-BDP Model 1 | 45.30% | 77.57% | 93.94% | 97.89% |
| DNN-BDP Model 2 | 48.80% | 82.16% | 97.23% | 99.89% |
| DNN-BDP Model 3 | 48.76% | 82.04% | 97.41% | 99.91% |
| DNN-BDP Model 4 | 50.82% | 79.37% | 94.20% | 99.91% |
| DNN-BDP Model 5 | 34.38% | 65.57% | 86.15% | 95.91% |
| DNN-BDP Model 6 | 25.99% | 59.11% | 88.12% | 97.76% |
| DNN-BDP Model 7 | 15.67% | 21.66% | 30.81% | 45.85% |
| DNN-BDP Model 8 | 12.17% | 18.85% | 23.66% | 30.82% |
| DNN-BDP Model 9 | 56.39% | 88.87% | 96.67% | 97.07% |
| DNN-BDP Model 10 | 58.36% | 91.56% | 99.36% | 99.99% |

## C. Overall Performance of HDL-BDQ Model

Since the superior performance of Model 10 for DNN-BDP, we have selected Model 5 from DNN-UBDF and Model 10 from DNN-BDP as the two sequential DNN models for the proposed HDL-BDQ model. The overall performance of HDL-BDQ model is evaluated by connecting DNN-UBDF and DNN-BDP together. Table VII present the overall performance of the HDL-BDQ model and the benchmark models. At the 15% error tolerance level, the proposed HDL-BDQ model achieves an accuracy of 91.7%, whereas the NNBD model only attains 83.1%. It is evident that the proposed HDL-BDQ model can outperform the benchmark models completely across all tolerance levels. The proposed HDL-BDQ model achieves a high accuracy especially when the tolerance is in 5% and 10%. Note that the results for three models shown in Table VII are based on the same training dataset that do not undergo any filtering or pre-processing.

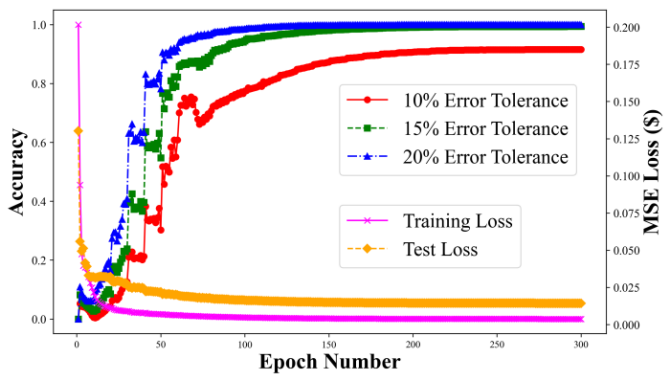

Fig. 6. Training result of Model 10 for DNN-BDP.

Table VII Performance Comparison of Battery Degradation Prediction

| Error Tolerance | 5% | 10% | 15% | 20% |
|---|---|---|---|---|
| HDL-BDQ | 37.4% | 73.4% | 91.7% | 97.3% |
| NNBD | 31.4% | 57.0% | 83.1% | 97.3% |
| NNBD2 | 27.3% | 55.6% | 79.2% | 91.4% |

Table VIII Microgrid Testbed

| Main Components | Diesel Generator | Wind Turbines | Solar Panels | Lithium-ion BESS |
|---|---|---|---|---|
| Size | 180kW | 1000kW | 1500kW | 300kWh |

## D. Microgrid Testbed

A typical grid-connected microgrid integrated with renewable energy sources is demonstrated as a test platform to evaluate the performance of the proposed HDL-BDQ embedded LAS model and LOD algorithm. The testbed comprises diesel generator, wind turbines, residential homes equipped with solar panels, and lithium-ion BESS with a 90% charging/discharging efficiency, which also accounts for power electronics losses [46]. The parameters for the main components are listed in Table VIII. The load data is based on 1000 residential houses and the ambient temperature and available solar power for a 24-hour period are sourced from the Pecan Street Dataport [47]. The wholesale electricity price is utilized based on locational marginal price (LMP) of day-ahead market at Austin, Texas. These LMP profiles are obtained from ERCOT [48]. The LAS optimizing problem was solved on a computer with an AMD® Ryzen 7 3800X, 32 GB RAM, and Nvidia Quadro RTX 2700 super (8 GB GPU). The 24-hour LAS problem was solved using the Pyomo [49] package and the Gurobi solver [50]. The neural network is built with the PyTorch package [51].

## E. Solving Results of HDL-BDQ Embedded LAS Problem

In this section, the results of HDL-BDQ embedded LAS problem that are obtained with the LOD algorithm as well as the results of the benchmark models are presented in Table IX. It can be easily observed that the total cost obtained from the LOD algorithm is the lowest among the three models. The traditional LAS model is the most efficient model, but it has the highest total cost and degradation cost compared to the other models. Note that the degradation cost of the traditional model is retrieved by passing the solution to the HDL-BDQ model, which also represents the highest degradation cost as battery degradation is not considered at all in the traditional model. Furthermore, the degradation cost of the linear BDC model concerning the Traditional one decreases by 58%, while the total cost is reduced by 1.25%, respectively. On the other hand, LOD algorithm can reduce the degradation cost by 72.8% and the total cost by 3.47%. Moreover, the operation cost of LOD algorithm does not increase significantly. Although the solving time is the highest for LOD algorithm, it is still acceptable as a trade off in a LAS problem, and thus, we can conclude that the proposed LOD algorithm outperforms the other two benchmark models.

Table IX Results for proposed strategies

| LAS | LOD | Traditional | Linear BDC |
|---|---|---|---|
| Total Cost ($) | 493.57 | 511.3 | 504.92 |
| Operation Cost ($) | 483.65 | 474.77 | 489.6 |
| Degradation Cost ($) | 9.92 | 36.53 | 15.32 |
| Solving time (s) | 5.76 | 0.34 | 0.42 |
| Iteration Numbers | 32 | N/A | N/A |

Sensitivity tests have been conducted on the $\alpha$ value to assess its impact on the performance of the LOD algorithm when solving LAS problems. The results from Table X demonstrate that a larger alpha value can achieve the lowest total cost in fewer iterations and less solving time. However, as a tradeoff, the optimal total cost with a larger alpha value may be slightly higher. A smaller alpha value results in a smaller area being cut from the feasible solution area in each iteration, which requires more iterations to converge to the optimal solution. However, the optimal solutions for different alpha values do not show significant differences. Therefore, a higher value of alpha, such as 0.05, is preferred due to its high computing efficiency.

Table X Results of sensitivity analysis with different α values

| Alpha | Number of Iterations | Optimal Total Cost ($) | Time (s) |
|---|---|---|---|
| 0.01 | 110 | 494.27 | 29.15 |
| 0.02 | 57 | 494.27 | 14.82 |
| 0.03 | 36 | 493.37 | 9.65 |
| 0.05 | 22 | 493.57 | 5.76 |
| 0.1 | 11 | 493.80 | 3.16 |
| 0.2 | 5 | 494.13 | 1.77 |

The detailed results of the iterative process of LOD algorithm are shown in Fig. 7, including the degradation cost, operation cost, and total cost. The capital investment cost of BESS $c_{BESS}^{Capital}$ is set at $400 per kWh, the salvage value $c_{BESS}^{SV}$ is set at $0 per kWh, and the value of α is set to 0.05. From the figure, it can be observed that proposed LOD algorithm finds the lowest total cost at the 23rd iteration, with a total cost of $493.57, including a degradation cost of $9.92 and a microgrid operation cost of $483.65. This result demonstrates effectiveness of the proposed LOD algorithm in determining the optimal solution. The stopping criterion was not applied in this instance to demonstrate the system's behavior when the battery usage is further limited until idle. In normal circumstances, the iteration would have stopped at the 33rd iteration as the minimum total cost was found at the 23rd iteration. Although the battery degradation cost continues to decrease after the 23rd iteration, the overall cost begins to rise as a result of the steep incline of microgrid operation cost.

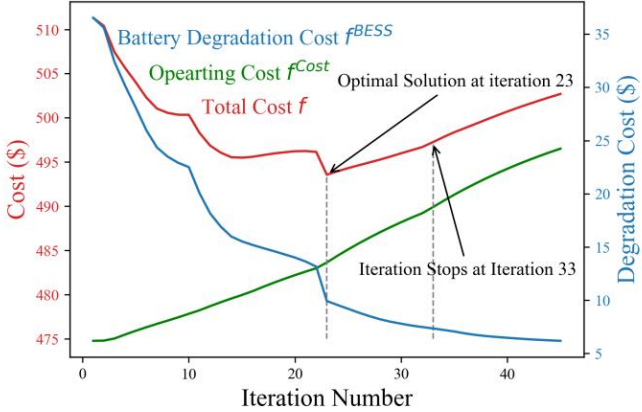

Fig. 7. Results of the LOD algorithm.

Fig. 8 shows the scheduled BESS operation for different models, where a positive output represents discharging mode and negative output represents charging mode. The scheduled BESS operations for the traditional LAS model, Linear BDC model and the proposed HDL-BDQ embedded LAS model are all shown in Fig. 8. It can be observed that for the traditional LAS that does not consider the battery degradation, BESS operates at a wider output range from -150 kW to 150 kW in seven different time intervals. When battery degradation is considered in the Linear BDC model, BESS is scheduled to charge and discharge in a narrower range and in four active time intervals. In the HDL-BDQ embedded LAS model, BESS is scheduled to operate only in three active time intervals and in a narrower power range. The BESS operation patterns for different models are consistent in most of the time intervals, which proves the effectiveness of those three models.

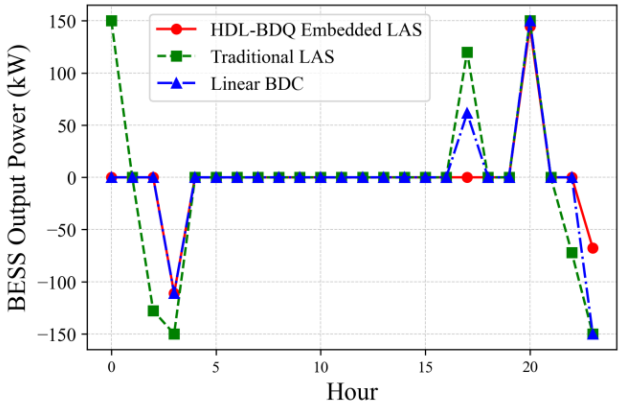

Fig. 8. BESS scheduled operations comparison.

A two-week LAS problem has been tested with a microgrid testbed to evaluate the performance of the proposed LOD algorithm in a longer look ahead schedule. The results of the two-week look-ahead scheduling (LAS) problem shown in Table XI demonstrate the effectiveness of our LOD algorithm in finding the optimal solution while considering battery degradation. It achieved a 60.8% reduction in battery degradation compared to the traditional LAS problem, with a 0.69% reduction in total cost. Although this reduction in total cost is not as significant as the results from the 24-hour look-ahead, which is based on a shorter timeframe, it is still notable. In contrast, the linear BDC model performs poor in this scenario. The total cost increases even when the battery degradation cost is considered in the LAS problem. While the decrease in the degradation cost indicates some effectiveness of the linear BDC model, the fact that the total cost does not decrease despite the decrease in degradation cost suggests that the solution has deviated from the optimal path.

Table XI Results for proposed strategies with two week look ahead.

| LAS | LOD | Traditional | Linear BDC |
|---|---|---|---|
| Total Cost ($) | 6923.7 | 6971.8 | 6969.7 |
| Operation Cost ($) | 6783.3 | 6614.0 | 6625.4 |
| Degradation Cost ($) | 140.4 | 357.8 | 344.3 |
| Solving time (s) | 197 | 5.89 | 8.4 |
| Iteration Numbers | 31 | N/A | N/A |

An extensive test case has been conducted for a one-year duration using the HDL-BDQ embedded LAS framework. We selected 12 representative days to accurately reflect each month throughout the year. The data was sourced and scaled based on ERCOT's 2023 data [48]. As detailed in Table XII, the results indicate that the model's performance over the one-year period aligns with the effectiveness observed in the daily and bi-weekly LAS scenarios. In this year-long testing of the LAS model, there is a significant reduction in total costs and battery degradation costs compared to the benchmark models. Specifically, the LAS model achieves a total cost of $302,069, which represents a 2.33% reduction compared to the Traditional model ($309,275) and a 0.56% reduction compared to the Linear BDC model ($303,759). In terms of battery degradation costs, the LOD algorithm incurs only $2,871, marking an 81.38% reduction compared to the Traditional model's $15,422 and a 61.18% reduction compared to the Linear BDC model's $7,396. These reductions underscore the LAS model's effectiveness in optimizing cost efficiency and minimizing battery wear over the long term, making it a more sustainable and economically viable option.

Table XII Results for proposed strategies with yearly look ahead.

| LAS | LOD | Traditional | Linear BDC |
|---|---|---|---|
| Total Cost ($) | 302,069 | 309,275 | 303,759 |
| Operation Cost ($) | 299,198 | 293,853 | 296,363 |
| Degradation Cost ($) | 2,871 | 15,422 | 7,396 |
| Solving time (s) | 2,790 | 129 | 120 |

To evaluate the resilience of the proposed algorithm, tests were conducted using an IEEE-24 bus system based on [52], as illustrated in Fig. 9, with the results presented in Table XII. The BESS is located at bus 7 with a capacity of 200MWh. The total cost is calculated as the sum of the operation cost (generators' fuel cost) and the equivalent battery degradation cost. Since the traditional model does not consider battery degradation, the equivalent battery degradation cost for the traditional model in Table XII was estimated by extracting the BESS scheduling profile from the traditional LAS problem and using it as input for the HDL-BDQ model. The results align with expectations, showing a reduction in total cost achieved by the proposed LOD algorithm. Specifically, the total cost decreased by 1.26% compared to the traditional model, with a significant 40% decrease in battery degradation cost. Although the operating cost increased by 0.84% due to changes in BESS schedules, this increment was minor compared to the reduction in battery degradation cost.

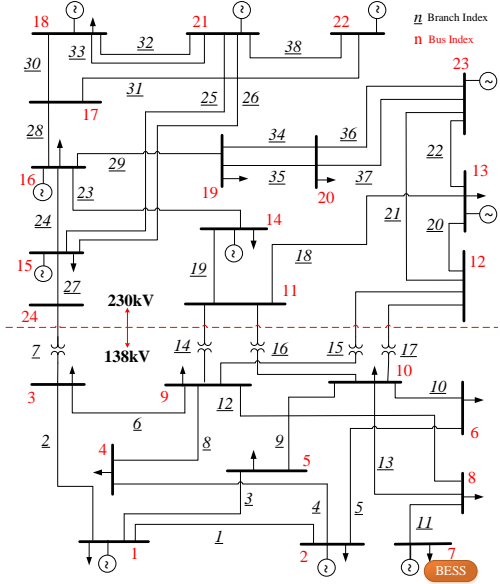

Fig. 9. IEEE-24 Bus testbed.

Table XII Results for proposed strategies with IEEE 24 bus system.

| LAS | Operating Cost ($) | Battery Degradation Cost ($) | Total Cost ($) |
|---|---|---|---|
| Traditional | 296,799 | 16,104 | 312,903 |
| LOD | 299,298 | 9,662 | 308,960 |
| Reduction | -0.84% | 40% | 1.26% |

## VI. CONCLUSIONS

In this paper, a hierarchical deep learning-based battery degradation quantification model is proposed to predict the BESS degradation value for each scheduling period. Multiple potential DNN models are developed and tested separately to determine the most accurate HDL-BDQ model that is the most effective combination of the DNN-UBDF and DNN-BDP models. The LOD algorithm is developed to solve the HDL-BDQ embedded look-ahead scheduling problem that is hard to solve directly due to the highly non-linear characteristics of the proposed HDL-BDQ model. The proposed LOD algorithm can solve the HDL-BDQ embedded look-ahead scheduling optimization problem and calculate the battery degradation cost iteratively and effectively to determine the optimal solution with the lowest total cost.

The simulation results demonstrate that the proposed HDL-BDQ is more advanced than the single stage NNBD model since it requires less training data and achieves higher training accuracy (91.7%). The proposed various potential architectures for the two sequential DNN models also help determine that the IR and ELCN are the most correlated battery degradation parameters. Furthermore, the proposed LOD algorithm can successfully obtain the optimal solution efficiently in a microgrid testbed. The total cost is reduced by 3.47% and the degradation cost is reduced by 72.8% compared to the traditional look-ahead scheduling model. The case studies in this paper demonstrate the performance of the proposed HDL-BDQ model in a microgrid application. Moreover, the proposed HDL-BDQ model and LOD algorithm can also be applied in other energy systems that contain BESS.

## VII. FUTURE WORK

The training dataset for the proposed model is currently based on simulated data, which limits the robustness of the HDL-BDQ model. Despite this, the HDL-BDQ model presents a well-defined data-driven framework for battery degradation. As realistic battery aging data becomes available, we plan to incorporate this data into the model to refine and enhance its performance. A key strategy we intend to pursue is leveraging transfer learning with the trained HDL-BDQ model to adapt it for different battery types, thus allowing us to develop battery degradation models with significantly less data. Furthermore, we aim to implement more systematic optimization techniques, such as grid search or Bayesian optimization, to further enhance model's performance. We also plan to develop and enhance the capability of the HDL-BDQ model to address long-term LAS problems more effectively. This will involve adapting the model to better handle extended time horizons and improve its performance and efficiency for annual or seasonal scheduling scenarios.

IEEE Transactions on Smart Grid 12